
\voffset=-0.6in
\hbadness=10000
\vbadness=10000
\baselineskip=24pt
\magnification=1200
\line {May 1993 \hfil  INFN-CA-93-12}
 \vfill
\centerline {\bf STRING QUANTUM SYMMETRIES  AND THE SL(2,${Z\kern -4.6pt Z}$)
 GROUP }
\vskip 1.2in
\centerline {Mariano Cadoni}
\vskip .2in
\centerline {Istituto Nazionale di Fisica Nucleare, Sezione di Cagliari}

\centerline {Via Ada Negri 18, I-09127 Cagliari, Italy and Dipartimento}
\centerline {di Fisica, Universit\`a di Cagliari}

\vfill
\centerline {\bf ABSTRACT}
We  prove,  using arguments relying only on  the "special K\"ahler" structure
of
the moduli space of the Calabi-Yau three-fold,  that in the case of one single
modulus the quantum modular group of the string effective action corresponding
to  Calabi-Yau  vacua can not be SL(2,${Z\kern -4.6pt Z}$).

\eject

A lot of effort has been recently devoted in order to understand the local
and global properties of the moduli space of the Calabi-Yau manifolds [1-3].
Some of the low-energy parameters, corresponding to the vacua of the heterotic
string described  by compactifications on a Calabi-Yau three-fold
or equivalently by a  N=2, c=9 world-sheet
Superconformal Field Theory (SCFT),  have been computed by using techniques
 of algebraic geometry$\,$[2]. The relevant information about these couplings
is
encoded in a 4th-order linear holomorphic differential equation, particular
case
of the so called "Picard-Fuchs" equations that can be written for any
Calabi-Yau d-fold [4]. It was also realised that the existence of such
differential $\,$ equations implies the geometrical structure of the moduli
space to be of restricted type,  namely "special K\"ahler" [1,6].
A crucial role in this contest is played by a group of discrete symmetries,
 the so called modular (duality) group,  which acts isometrically on the
K\"ahler
 manifold. Owing to their "stringy" origin they are believed to be preserved by
non perturbative world-sheet effects, consequently they have been used
to constrain the form of the low-energy couplings [7,8].
Despite a lot of effort in trying to understand the meaning of these
symmetries we are still far away of having full control of them.
What seems however clear is that the modular group related to toroidal
and orbifolds compactifications,  namely SL(2,${Z\kern -4.6pt Z}$),
has not a big chance
to survive as the right "quantum" modular group of the string.
In fact for the case of the exactly soluble SCFT given by the pair of
 Calabi-Yau manifolds of Candelas, the modular group was shown to be not
 SL(2,${Z\kern -4.6pt Z}$)
but a different discrete subgroup of SL(2,$\rm I\!\rm R$) [2].
Nevertheless nothing in principle prevents that some specific Calabi-Yau
three-fold could exist whose modular group is exactly SL(2,${Z\kern -4.6pt
Z}$).
In this letter we will rule out this possibility  for the one modulus case
by showing that  the only choices for the Yukawa coupling,  compatible
with the SL(2,${Z\kern -4.6pt Z}$) symmetry,  are either
$W=const\not= 0$ \footnote{$^1$}{We use $W$ for the Yukawa couplings. The
reader
should avoid any confusion with the standard supergravity notation where $W$
denotes the superpotential.},
corresponding to toroidal compactifications i.e the large
radius limit of Calabi-Yau compactifications,  or $W\sim \eta^8$     ($\eta$
is the Dedekind function).
The second possibility is of no direct physical interest, at least for the one
modulus case,  because gives $W=0$ in the large radius limit.

The moduli space of Calabi-Yau manifolds  is locally a special K\"ahler
manifold defined by the K\"ahler potential [1,6] (we will only consider the one
modulus case)
$$K=-\ln(V(iQ)V^{\dag}),\eqno(1)$$
where Q is the symplectic metric
$$ Q= \pmatrix{0&1\!\!1_2\cr -1\!\!1_2&0\cr},\eqno(2)$$
and
$$V=\left( X^A( z),F_A( z)\right):=\left(X^0,X^1,F_0,F_1\right),
\quad F_A={\partial F\over \partial X^A}.\eqno (3)$$
$X^A$ and $F$ are both holomorphic and $F(X^A)$ is homogeneous of degree
two in $X^A$.
The K\"ahler potential of eq. (1)  is manifestly invariant under Sp(4,
 $\rm I\!\rm R$)
 global
transformations.
In the following we will consider "special coordinates" defined by
$$t={X^1\over X^0} ,\qquad X^0=1. \eqno (4)$$
In these coordinates the Yukawa coupling $W$ is given in terms of  $F$ by
$$W=\partial^3 F .\eqno(5)$$
All special K\"ahler geometries in one dimension lead to a 4th-order
differential equation characterised by some coefficients $\omega_2$ and
$\omega_4$ [5]
$$\left[\partial^4+\omega_2\partial^2 +\omega_2^\prime \partial +{3\over10}
\omega_2^{\prime\prime}+{9\over 100}\omega_2^2 +\omega_4\right]V=0.\eqno(6)$$
Under coordinates change $\tilde z=\tilde z( z)$,  $\omega_2$,  $\omega_4$
and $V$ transform as
$$\eqalign{\tilde\omega_2 &= \xi^{-2}\left[\omega _2-5 \left\{\tilde z ,
 z\right\}
\right],\cr \tilde \omega_4&= \xi^{-4}\omega_4,\cr
 \tilde V&= \xi^{3\over2}V,\cr}\eqno(7)$$
where $\xi =\partial \tilde z/\partial z$ and
 $\left\{\tilde z, z\right\}$
is the schwarzian derivative.

The modular group associated to the special K\"ahler manifold is the subgroup
$\Gamma$ of Sp(4,${Z\kern -4.6pt Z}$) which acts isometrically
$$K\left(\Gamma  z\right)=K( z) -\Lambda( z)-\Lambda(\bar z).
\eqno(8)$$
The key point is that in terms of the 4th-order differential equation
imposing the condition (8) amounts to require that under the action
of $\Gamma$,
$\omega_2$ and $\omega_4$ are form-invariant, i.e
$$\tilde\omega_2(\tilde z)=\omega_2(\tilde z),\quad \tilde\omega_4(\tilde z)
=\omega_4(\tilde z).\eqno(9)$$
The previous equations impose severe constraints on the form of $\omega_2$
and $\omega_4$ and eventually enable one to write them explicitly.
Let us now consider $\Gamma$=$\,$SL(2,${Z\kern -4.6pt Z}$) and assume
 that in terms of the
special coordinate $ t$ it is realised as the linear fractional transformation
(the general case will be discussed later on this paper)
$$\Gamma :\quad \tilde t ={a t+b\over ct+d},\quad ad-bc=1, \quad a,b,c,d
 \in {Z\kern -4.6pt Z}.\eqno(10)$$
{}From the first two eqs. in (7) it follows that for $\Gamma$ given by
(10) eqs. (9) admit only the two solutions
$$\omega_2=\omega_4=0,\eqno(11)$$
$$\omega_2(\tilde t)=\left(ct+d\right)^4\omega_2(t), \qquad \omega_4(\tilde t)=
\left(ct+d\right)^8\omega_4(t).\eqno(12)$$
In order to discuss the previous solutions we need the expressions relating
$\omega_2$ and $\omega_4$ to the Yukawa coupling $W$.
They can be found in Ref. [5]. We have
$$\omega_2 = {1\over 2W^2} \left(4W W^{\prime\prime}-5 W^{\prime 2}\right),$$
$$\omega_4 ={1\over 100 W^4}\left(175 W^{\prime 4}-280 WW^{\prime 2}W^{
\prime\prime}+49W^{2}W^{\prime\prime2} +70W^2W^{\prime }W^{\prime\prime\prime}
-10W^{3}W^{\prime\prime\prime\prime}\right).$$
 To put  these equations in a more manageable form we introduce a new
variable $U$ related to $W$ by $W=U^{-4}$,  in terms of which we get
$$\omega_2 =-8\, {U^{\prime\prime}\over U}\eqno(13)$$
$$\omega_4 =- \left(U^\prime\over U\right)^2\omega_2 -{1\over 2}
 \left(U^\prime\over U\right)
\omega_2^\prime -{1\over 20} \,\omega_2^{\prime\prime} +{7\over 200}\,
 \omega_2^2.\eqno(14)$$

Let us now discuss the two solutions (11) and (12).
The solution (11) corresponds to $ W=const$ or equivalently to cubic $F$
functions which describe toroidal compactifications (the large radius
limit of Calabi-Yau compactifications).
The differential equation (6) is solved by [5] (Our solution differs
from that of Ref. [5] owing to the choice of the symplectic
metric in (2))$$ V=\left( 1,\,t,\,-{1\over 6} t^3,\,{1\over 2}
t^2\right).\eqno(15)$$
Note that in this case SL(2,${Z\kern -4.6pt Z}$) appears as a discrete subgroup
of
Sp(4,$\rm I\!\rm R$).
In fact from the action of $ M\in$ Sp(4,$\rm I\!\rm R$) on V
$$\tilde V=VM,\eqno(16)$$and from (10) we get the embedding of
 SL(2,${Z\kern -4.6pt Z}$) in
Sp(4,$\rm I\!\rm R$)
for the case under consideration\footnote{$^2$}{This matrix was first given in
Ref. [11]}
$$M=\pmatrix{d^3&bd^2&-b^3/6&b^2d/2\cr 3cd^2&2bcd+ad^2&
-ab^2/2&b^2c/2+abd\cr -6c^3&-6ac^2&a^3&-3a^2c\cr
6c^2d&2bc^2+4acd&-a^2b&2abc+a^2d}.\eqno(17)$$
By changing the symplectic basis one can achieve $M\in$ Sp(4,${Z\kern -4.6pt
Z}$).

We turn now to the solutions defined by (12).
Eq. (13) together with the first equation in (12) enables one to find how
 U and W transform under SL(2,${Z\kern -4.6pt Z}$). We get
$$U(\tilde t)=\left(ct+d\right)^{-1} U(t),\eqno(18)$$
$$W(\tilde t)=\left(ct+d\right)^4W(t).\eqno(19)$$
Eqs. (19) and (12) tell us that $\omega_2$, $\omega_4$
and $W$ have to transform  under SL(2,${Z\kern -4.6pt Z}$) as modular
 functions of definite weight.  They  should therefore be expressible in
B terms of modular forms of SL(2,${Z\kern -4.6pt Z}$).
We will concentrate in the following only on the Yukawa coupling $W$,
the coefficients $\omega_2$, $\omega_4$ being determined by
(13) and (14) once W is known.
The most natural way to implement eq. (19) is to set
$$W(t)=\eta^8(t),\eqno(20)$$
where $\eta$ is the Dedekind eta-function
$$\eta (t)=\exp\left({i\pi\over12}t\right) \prod _n\left[1-\exp\left(2i\pi
n t\right)\right].\eqno(21)$$
The solution of the differential equation (6) can be now written as
$$ V=\left( 1,t,F_0,F_1\right),\eqno(22)$$
with $ F_0$ and $F_1$ satisfying
$$\partial^2 F_0=-t\,\eta^8,\eqno(23)$$
$$\partial^2 F_1=\eta^8.\eqno(24)$$
Eqs. (20-24) describe a well defined solution to one-dimensional special
K\"ahler geometry with SL(2,${Z\kern -4.6pt Z}$) isometries. However, as one
can easily see
from the asymptotic behaviour of $\eta$,  as $t\rightarrow i\infty$,
$W\rightarrow 0$.
The large radius limit does not correspond  to the usual one $W=const$ for
the Calabi-Yau moduli space.
Naturally one could search for solutions of (19) different from (20).
The only possible generalizations of eq. (20) satisfying the modular
constraint (19) are either $W(t)=\eta^8(t) S(J)$ or $ W(t)= G_4(t) H(J)$,
where $G_4(t)$ is the Einsestein function of weight four and  $S(J)$, $ H(J)$
are rational functions of the absolute modular invariant $J(t)$. A  definition
of the modular forms  $G_4(t)$ and  $J(t)$ can be found in Ref. [8];
 for more details on the modular forms of  SL(2,${Z\kern -4.6pt Z}$) see e.g
 Ref. [9]. One can easily check,  using the asymptotic expansions of $\eta$,
$G_4$ and $J$,  that there is no choice for $S(J)$ or $H(J)$ which
in the large radius limit reproduces $W=const$.

To complete our analysis on the solutions defined by (12) we have to find
the embedding of SL(2,${Z\kern -4.6pt Z}$) in Sp(4,${Z\kern -4.6pt Z}$)
 which is consistent with (20-24).
To this purpose we write down explicitly the general Sp(4,${Z\kern -4.6pt Z}$)
 transformation (16) in special coordinates
$$M=\pmatrix{A&B\cr C&D,\cr }\eqno(25)$$
$${\tilde X^1\over\tilde X^0}= \tilde t ={{a_{12}+a_{22} t+c_{12}F_0+c_{22}F_1}
\over a_{11}+a_{21} t+c_{11}F_0+c_{21}F_1},\eqno(26)$$
$$\tilde F_0= b_{11} +b_{21}t+d_{11}F_0+d_{21}F_1,\eqno(27)$$
$$\tilde F_1= b_{12} +b_{22}t+d_{12}F_0+d_{22}F_1,\eqno(28)$$
where $a_{ij},b_{ij},c_{ij},d_{ij},\,\,  i,j=1,2$ are the matrix elements
of the matrices $ A,B,C,D$ respectively and we took
 $ M\in$ Sp(4,${Z\kern -4.6pt Z}$),  i.e the entries of $M$ are integers
 verifying $ M^TQM=M.$
The embedding of SL(2,${Z\kern -4.6pt Z}$) in Sp(4,${Z\kern -4.6pt Z}$) is now
easily found setting in (25)
$$B=C=0,\quad A=\pmatrix{d&b\cr c&a\cr },\quad D={(A^T)}^{-1},\quad
 ad-bc=1.\eqno(29)$$
The transformation (26) reduces now to the SL(2,${Z\kern -4.6pt Z}$)-one
 given by (10) whereas (27) and (28) become
$$\tilde {\bf F}={\bf FD},\eqno(30)$$
where  the matrix $\bf D$ is  given as in (29) and ${\bf F}$ is the row
vector $F_A$.
On the other hand the transformation of $F_A$ leaving invariant the solutions
of (23-24) is found   using (10) and  (19):
$${\bf F}(\tilde t)=(ct+d)^{-1}{\bf F}(t){\bf D}.\eqno(31)$$
 The action of the modular group leaves the K\"ahler potential invariant up
to a K\"ahler transformation. This implies in general
$$\tilde {\bf F}(t)=u(t){\bf F}(\tilde t).$$
With the identification $u(t)=(ct+d)$ (30) and (31) are equivalent,
demonstrating
that the embedding (29) is the right one.
Our prove is now complete.  There is no solution to one dimensional
K\"ahler geometry exhibiting SL(2,${Z\kern -4.6pt Z}$) isometries and
 having as large radius
limit $W=const$,  except the limiting case itself.
Because any Calabi-Yau compactification should result in a particular
special K\"ahler geometry for the moduli space,  one can exclude,  in the
one-dimensional case,  SL(2,${Z\kern -4.6pt Z}$) as the "quantum" modular group
of
the string, at least if one wants to recover for large radii the usual field
theoretical limit.

Our argumentation is based  on the form (10) for the SL(2,${Z\kern -4.6pt Z}$)
transformations.  One can wonder if there is some  different realization
of the SL(2,${Z\kern -4.6pt Z}$) symmetry for which our prove does not hold.
This possibility can be ruled out by the following arguments.
The group Sp(4,$\rm I\!\rm R$) has only two independent SL(2,$\rm I\!\rm R$)
 subgroups.
We can identify the two realizations of the SL(2,${Z\kern -4.6pt Z}$) group
 defined  by (17) and (29) as coming from a discretisation of these two
SL(2,$\rm I\!\rm R$).
  There can not be any other independent SL(2,${Z\kern -4.6pt Z}$) $\subset$
 Sp(4,$\rm I\!\rm R$) because the existence of such a
subgroup not coming from a discretisation of an SL(2,$\rm I\!\rm R$)
 would imply that Sp(4,$\rm I\!\rm R$) splits in disconnected parts related
 exactly by this discrete subgroup.
We know that this is not the case.
As a consequence every realization of SL(2,${Z\kern -4.6pt Z}$)$\subset$
Sp(4,$\rm I\!\rm R$)
should be
related either to (17) or to (29) by a change of the symplectic basis,
i.e $M\rightarrow g^{-1}Mg$.
Because we should regard these realizations as equivalent,  we conclude
that (17) and (29) exhaust all the possible physical situations.

Let us end with some final comments. The solutions described by (20-24)
even though owing to their asymptotic behaviour do not seem to be
interesting in the one modulus case could play a role in the many moduli
case.  In this situation the behaviour $W\rightarrow 0$ as $t\rightarrow
 i\infty$ for some of the  moduli should not be rejected as unphysical.
The derivation of (20-24) was made exclusively on the ground of special
K\"ahler geometry. Hence we do not know if and how this has a correspondence
in terms of Calabi-Yau compactifications.
\break
In the above discussion we considered only Calabi-Yau compactifications.
However special K\"ahler geometry is a property of  general (2,2)
compactifications. For this general situation  the large radius behaviour of
the Yukawa couplings can not be used,  even in the one-modulus case,  to
discard
the solution (20);  in this context it might  be of physical relevance.
The final remark concerns the form (20) for the Yukawa coupling and gives
an intuitive explanation of the kind of obstruction one is faced with,
when trying to promote SL(2,${Z\kern -4.6pt Z}$) to  the "quantum" modular
group
of the string.
It has been stressed in the literature that the translation symmetry,  part
of the SL(2,${Z\kern -4.6pt Z}$) group, $t\rightarrow t+1,$ has a "stringy"
origin and
should therefore always appear as part of the modular group.
The general form of the Yukawa coupling required by this symmetry is [10]
$$W(t)=\sum_{n=0}^\infty d_n \exp{(2 \pi\imath n t)}.\eqno(32)$$
The presence of the constant term $d_0$ in the summation is crucial for getting
the large radius behaviour
$W=const$.
The extension of the translations to the full SL(2,${Z\kern -4.6pt Z}$) group
brings the
inversion
$t\rightarrow -1/ t$ into play.
Invariance under the inversion forbids the constant term $d_0$ and restricts
(32) to be proportional to the $\eta$ function as in (20),  with the
consequent ill behaviour in the large radius limit.

{\bf Acknowledgments}
\smallskip
I thank R. D'Auria for illuminating discussions and valuable comments.

\smallskip
\centerline {\bf References}
\smallskip

[1]  P. Candelas and X. C. de la Ossa, Nucl. Phys. {\bf B355} (1991) 455.

\hangindent=0,45in \hangafter=1
[2]  P. Candelas, X. C. de la Ossa, P. S. Greeen and L. Parkes, Phys. Lett
     {\bf B258} (1991)118;
     P. Candelas, X. C. de la Ossa, P. S. Greeen and L. Parkes,
    Nucl. Phys. {\bf B359} (1991) 21.

\hangindent=0,45in \hangafter=1
[3] P. Candelas, M. Lynker and R. Schimmrigk, Nucl. Phys. {\bf B341} (1990)
383.

\hangindent=0,4in \hangafter=1
[4] W. Lerche,  D. J. Smit,  N. P. Warner,  Nucl. Phys. {\bf B372} (1992) 87;
    A. Cadavic and S. Ferrara,  Phys. Lett. {\bf B267} (1991) 193;
    S. Ferrara, and J. Luis, Phys. Lett. {\bf B278} (1992) 240.

\hangindent=0,4in \hangafter=1
[5] A. Ceresole, R. D'Auria,  S. Ferrara, W. Lerche and J. Louis,
    Int. Jou. Mod. Phys. {\bf A8} (1993) 79.

\hangindent=0,4in \hangafter=1
[6]  L. Castellani, R. D'Auria and S. Ferrara,  Phys. Lett. {\bf B241}
(1990) 57;
R. D'Auria,  S. Ferrara and P. Fr\'e, Nucl. Phys. {\bf B359} (1991) 705;
B. de Wit and A. Van Proeyen, Nucl. Phys. {\bf B245} (1984) 89; L.J. Dixon,
V.S. Kaplunovsky and J. Louis,  Nucl. Phys. {\bf B329} (1990) 27.

\hangindent=0,4in \hangafter=1
[7] For a review see S. Ferrara and S. Theisen, in Proc. of the Hellenic
Summer School (1989), World Scientific; A. Font, L.E Ib\'a\~nez, D.L\"ust
and F. Quevedo,  Phys. Lett. {\bf B245} (1990) 401;  P. Bin\'etruy,  and
M. K. Gaillard,  Nucl. Phys. {\bf B358} (1991) 121;
J. Schwarz, Caltech Preprints CALT-68-1728 (1991) and  CALT-68-1740 (1991);
M. Cveti\v c, A. Font, L. E. Iba\~ nez,  D. L\"ust, F. Quevedo
Nucl. Phys. {\bf B361} (1991) 194.

\hangindent=0,4in \hangafter=1
[8] S. Ferrara,  D. L\"ust, A. Shapere and S. Theisen,  Phys. Lett.
 {\bf B255} (1989) 363.

\hangindent=0,4in \hangafter=1
[9] B. Rankin, Modular  forms and functions, Cambridge U.P., Cambridge 1977.

\hangindent=0,4in \hangafter=1
[10] R. D'Auria and  S. Ferrara,   CERN Preprint, CERN-TH. 6777/93;
     A. Klemm,  S. Theisen Preprint LMU-TPW 93-08

\hangindent=0,4in \hangafter=1
[11] S. Ferrara, D. L\" ust, S. Theisen, Phys. Lett {\bf B242} (1990) 39.

\end